\documentclass[a4paper]{jpconf}
\usepackage{graphicx}
\usepackage{amsmath,amssymb}

\newcommand{\path}{./figs/}

\begin{document}
\title{Charm decay as a source of multistrange hadrons}

\author{M~Petr\'{a}\v{n}$^1$, J~Letessier$^2$, V~Petr\'{a}\v{c}ek$^3$ and J~Rafelski$^1$}

\address{$^1$ Department of Physics, University of Arizona, Tucson, AZ 85719, USA}
\address{$^2$ Laboratoire de Physique Th\'{e}orique et Hautes Energies, Universit\'{e} Paris 6,  F-75005}
\address{$^3$ Czech Technical University in Prague, Brehova 7, 115 19 Praha 1, Czech Republic}


\begin{abstract}
We describe a newly formulated approach to account for charm production and decay in Statistical Hadronization approach. Considering Pb--Pb collisions at $\sqrt{s_{NN}} = 2.76$ TeV at LHC we show that charm  hadron decays can be a significant contributor to the multistrange hadron abundance. We discuss the magnitude of expected effects as a function of charm yield.  
\end{abstract}

\section{Charm Production and Decay in Heavy Ion Collisions}
We demonstrate that in a precise analysis of hadron abundances within, e.g., the statistical hadronization model (SHM), it is necessary at LHC to allow for contribution to hadron yields from charm decays.  In order to accomplish this we need to know: 1) total charm quark yield in QGP at hadronization; 2) charm  hadronization pattern into hadrons; and 3) charm hadron decay distribution. 

Initial state charm production can be estimated by folding the $pp$ cross section with the number of initial nucleon-nucleon collisions. To be specific, we recall that in central Au--Au collisions at RHIC, the total charm cross--section has been measured to be $d\sigma_{c\bar{c}}/dy|_{y=0}=175\pm12\pm23\mu b$, and shown to scale with the number of binary collisions~\cite{Tlusty:2012ix}. Thus, in  most central  0\%--5\% centrality bin RHIC reactions, one finds $N_{cc}^\mathrm{RHIC}\equiv dN_{c+\bar{c}}/dy\simeq 4.3 $ for $y\in (-0.5,0.5)$ measured in the CM frame.

For the LHC energy in Pb--Pb collisions at $\sqrt{s_{NN}} = 2.76$ TeV at LHC,  theoretical considerations lead to $N_{cc}=246\pm 154$ ($123\pm77$ pairs), where the error is due to uncertainty in charm production cross section~\cite{Nelson:2012bc}.  It is believed that most of these charm pairs survive the parton thermalization and the sequel QGP evolution  to the hadronization stage. However, if the prediction of such high abundance is true, one must expect some significant charm reannihilation. This can  be checked in near the future within our kinetic flavor evolution model~\cite{Letessier:2006wn}. 

At QGP hadronization, all surviving charm hadronizes, producing heavy particles with hidden, open single charm, and multiple charm content. After traveling at most a few hundred $\mu m$, all these particles have decayed. For nearly all decay channels, it is practically impossible to tell which final state hadrons originate in a charm decay. This means that the hadron yields one measures contain input from post-hadronization charm  decay. If all particles were fed by charm decay with a strength that is similar to the SHM model production, the contribution of charm would simply be a shift in overall normalization. The principal objective we pursue here is to show that  this is not the case. We show that charm and strangeness are strongly correlated, and thus charm replaces some of strangeness production in QGP. Given $N_{cc}$, we predict yields of all charmed hadrons using the  statistical hadronization method.

\section{SHARE with CHARM}
We have developed a new program predicting the contribution of charm hadron decays and allowing a fit with charm feed accounted  into all hadron abundance. This program  comprises a CHARM module which adds charm decay hadron multiplicity into SHARE, the statistical hadronization model implementation we use~\cite{Torrieri:2004zz,Torrieri:2006xi}. The new SHARE with CHARM utility uses $N_{cc}$ as an additional fit parameter when analyzing hadron production in heavy--ion collisions. SHARE with CHARM further allows the differentiation of the temperature of charm hadronization $T_c$ from the temperature $T$ of other hadron production. However, for reason of space we refrain here from showing results with this refinement.

\begin{figure}[t]
\centering
\begin{minipage}[t]{0.44\columnwidth}
\includegraphics[height=50mm]{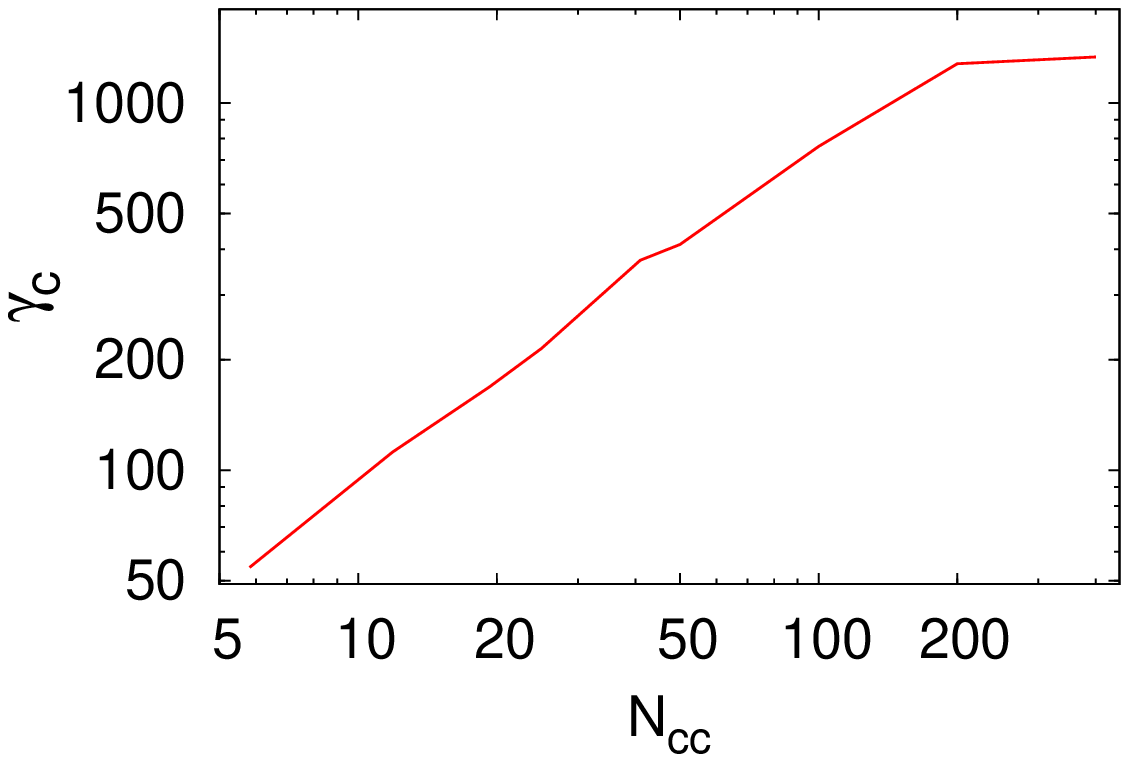}
\caption{\label{fig:gamc} The charm phase space occupancy as a function of total charm present at hadronization.}
\end{minipage}\hspace*{0.03\columnwidth}%
\begin{minipage}[t]{0.5\columnwidth}
\hspace*{-0.3cm}\includegraphics[height=50mm]{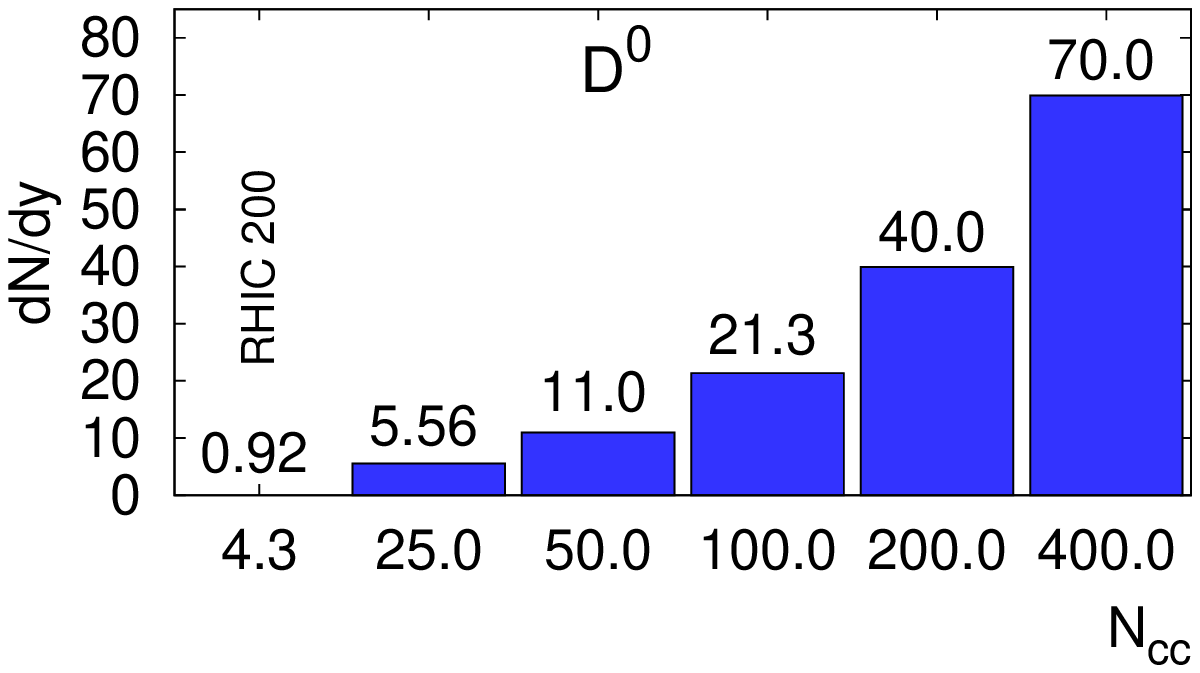}
\caption{\label{fig:dzero} The $D^0$ meson yield as a function of total charm abundance $N_{cc}$ at hadronization.}
\end{minipage}\hspace*{0.04\columnwidth}%
\end{figure}

SHARE with CHARM functions as follows: for a given  $N_{cc}$ in the first step we compute the distribution of charm among all charm hadrons following the principles of statistical hadronization at a prescribed temperature $T_c$.  To describe the normalization of charm yield, the phase space occupancy $\gamma_c$ has been introduced. Given $N_{cc}$, $\gamma_c$ is solved for using methods described in~\cite{Kuznetsova:2006bh}. Charm production  is not thermal, but is due to the initial hard parton scattering. Therefore it is expected to be strongly out of chemical equilibrium. In Figure \ref{fig:gamc}  we see the magnitude of  $\gamma_c$  as a function of  $N_{cc}$ for hadronization temperature $T_c=T$. We note that for $N_{cc}>50$, the magnitude of $\gamma_c>500 $ implies that some of the charm produced will reannihilate. This effect may explain how the large initial predicted yield can be consistent with the more realistic yield obtained in direct experimental effort, as we now describe.

Figure~\ref{fig:dzero} shows the expected  $dN_{D^0}/dy$ yield (number  above each column) for a prescribed value of  $N_{cc}$. Charmed hadrons in general show dependence on $N_{cc}$ proportional to the number of constituent (anti-)charm quarks with small variations due to changes in statistical parameters such as  $T$ and $\gamma_s$ arising when a fit is performed. The ALICE experiment at LHC has the capability of recognizing the charm decay in specific channels and thus can measure the charmed hadron yields. The  $D^0$ meson $p_\perp$--spectrum has been presented~\cite{ALICE:2012ab} for the 0-20\% centrality bin. Due to the uncertain low $p_\perp$ bin, the yield $dN/dy$ cannot be reliably integrated.  However, the other points allow  us to estimate by extrapolation to $p_\perp=0$ a yield $dN/dy|_{D^0}\in(1.3,9.0)$. 

Considering the results presented in  Figure~\ref{fig:dzero}, this implies  a 5 times larger total yield $N_{cc}\simeq 6-45$ charm and anti-charm quarks. Scaling from 0--20\% to the 0--5\% centrality by a  factor 1.35  we find $N_{cc}^\mathrm{LHC}\in (10,60)$ as the best direct experiment-based  estimate, much less than the central point of the initial production evaluation $N_{cc}=246\pm 154$~\cite{Nelson:2012bc}. Still, we choose to show expected yields for up to a value $N_{cc}=400$, and the yield of  $D^0$ mesons clearly is not compatible even with $N_{cc}=100$.

In the next SHARE with CHARM step, each of the produced charm hadrons is allowed to decay. Based on experimental decay data~\cite{Beringer:1900zz}, symmetry principles, and plausibility arguments, we prepared a complete charmed hadron decay table. This task took much effort, required to study of many additional theoretical references, and further improvements will certainly be necessary. One must see our present effort as a first step to be as complete  as  the current understanding of charm hadron decays.

In the final SHARE with CHARM step, the produced hadrons and their resonances are added to the SHARE initial SHM yields obtained for a set of chosen SHM parameters. Hadron resonances are  decayed; the total hadron yields are compared to the experimental data input field, and $\chi^2$ computed. An improved set of statistical parameters is proposed by MINUIT routine embedded in SHARE. The cycle of computations is carried out until SHARE concludes that the best local minimum of $\chi^2$ is achieved. 

\section{Charm decay feed}
We quantify the charm hadron decay contributions in the final hadron yields using previous analysis of central Au--Au collisions at RHIC~\cite{Rafelski:2004dp} and for LHC, we perform our study using the same input data as in our recent analysis of Pb--Pb collisions at LHC~\cite{Petran:2013qla,Petran:2013lja}. In order to quantify the expected effect in the absence of a measured charm hadron yield at LHC we perform a parametric study prescribing a value of  $N_{cc}\in (0,400)$ while fitting the yields of observed hadrons. Charm feed--down of light hadrons generally has a different pattern than the  light hadron production in QGP hadronization. Therefore, although the charm contribution may be hidden within the experimental errors of light hadrons, a very large abundance introduces a tension in the SHM fit of multiple hadron species with different strangeness content. 

\begin{figure}
\centering
\begin{minipage}[t]{0.47\columnwidth}
\includegraphics[height=45mm]{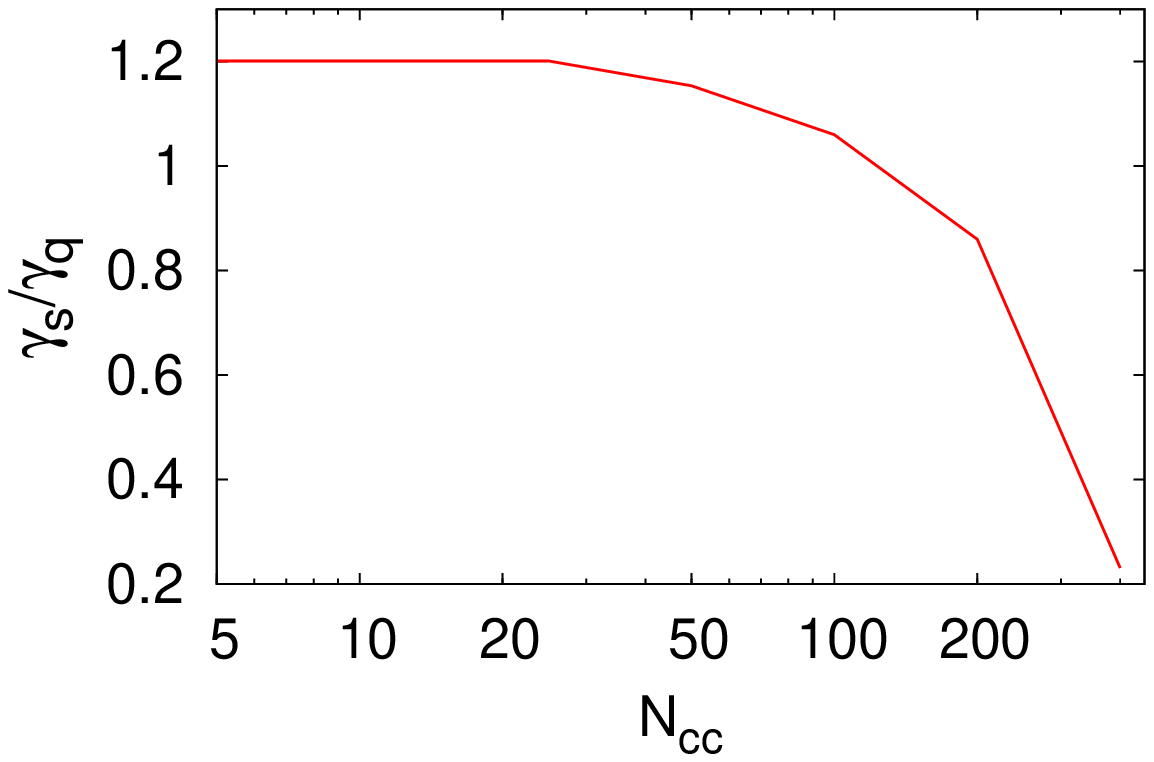}
\caption{\label{fig:gamsgamq} Strangeness over light quark phase space occupancy as a function of total charm present at hadronization.}
\end{minipage}\hspace*{0.04\columnwidth}%
\begin{minipage}[t]{0.47\columnwidth}
\hspace*{-0.2cm}\includegraphics[height=45mm]{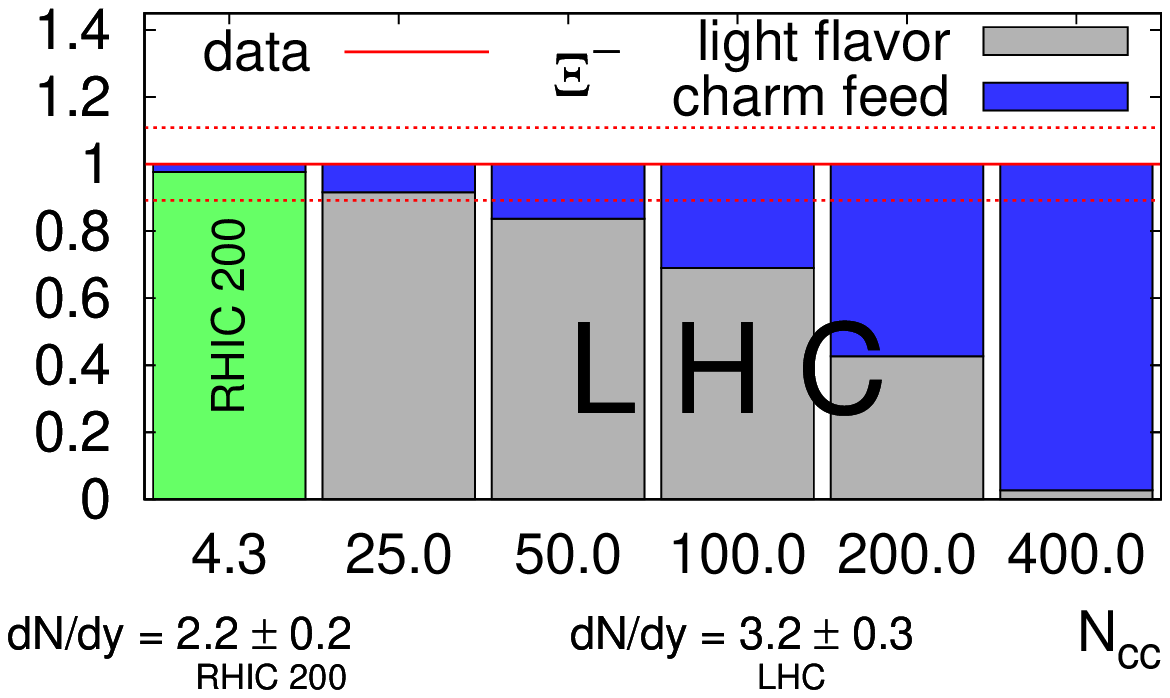}
\caption{\label{fig:ximinus} The relative contribution of charm decays to the yield of $\Xi^-$ as a function of charm abundance, see text for details.}
\end{minipage}
\end{figure}

In Figure \ref{fig:gamsgamq}, we show the change of the statistical parameter $\gamma_s/\gamma_q$, characterizing the  strangeness present at time of hadronization, resulting from the fit.  For the value  $N_{cc}=0$, the results of our prior fit apply~\cite{Petran:2013qla,Petran:2013lja}. There is a nominal change up to   $N_{cc}<50$. For $N_{cc}>50$ the charm decay feed replaces in a significant way the strangeness abundance, and for $N_{cc}=400$ strangeness in QGP at hadronization would be practically erased. Naturally, this is impossible, and it is clear that $N_{cc}>150$ is an unpalatable result also inconsistent with the $D^0$ spectral  results as we discussed.

During our study, we confirm a previous result~\cite{Kuznetsova:2006bh}, that charm decays contribute substantially to multistrange hadrons,  $\phi$, $\Xi$ and $\Omega$.  In Figure~\ref{fig:ximinus},  we show as an example the relative contribution to $\Xi^-$ yield as a function of total charm abundance $N_{cc}$ normalized to the experimental value indicated at the bottom for RHIC and LHC. The red dotted lines represent the experimental error, the blue part shows the contribution to $\Xi^-$ yield from charm decays. One can see that at RHIC-200, the charm contribution presents an insignificant part of the total yield. However, at LHC-2760, the contribution to $\Xi^-$ yield (and similarly other multistrange particles) exceeds the experimental error for  $N_{cc}>25$ (that is 13 or more pairs).

We further study the effect of charm  on the thermal parameters of the SHM fit. Volume $dV/dy$, chemical freeze-out (hadronization) temperature $T$ and light hadron phase space occupancy $\gamma_q$ show a very slow, but steady decrease within their respective errors. As discussed, charm decays are a significant source of strangeness and hence the strange phase space occupancy $\gamma_s$ decreases significantly upon the introduction of a relatively large charm yield. This is the cause why  multistrange particles  are affected the most by charm decays.

We investigated the effect of including the charm degree of freedom on the bulk properties of the particle source. We found that the physical bulk properties, such as energy density $\varepsilon$, entropy density $\sigma$, and pressure $P$ remain unchanged. Therefore, the charm contribution to particle yields does not affect our previous findings about universality of hadronization condition~\cite{Petran:2013qla}.

\section{Outlook}
Using the new SHM program, SHARE with CHARM, we found  that in RHIC Au--Au collisions at 200 GeV, the amount of charm has very little direct effect on hadron yields, whereas in LHC Pb--Pb collisions at 2.76 TeV, charm has an observable effect on  hadron yields. We have shown that the present best estimate of the total $c+\bar c$ yield $N_{cc}\simeq 250\pm 150$ seems too large compared to the reported $D^0$ spectrum. We have shown that   $N_{cc}\simeq 100$ (50 pairs) will contribute nearly 3sd to the yield of $\Xi$ and thus analysis of the hadron yields should include charm decays once such a yield is confirmed.

A precise yield of at least one of the charmed hadrons is necessary for an analysis of hadron yields with SHARE with CHARM to be possible.  Without charmed hadrons we find that a fit of $N_{cc}$ does not lead to a clear $\chi^2$ minimum: we find a nearly totally flat $\chi^2$ as a function of $N_{cc}$. This is of course good news in the sense that any acceptable in size measured charmed hadron yield will ``work''. If and when we have more than one charmed hadron yield measured directly, we can determine the difference in hadronization temperature $\delta T\equiv T_c-T$. With more than two charmed hadron yields available we begin to test  the hypothesis that statistical hadronization applies to charmed hadron production. We noted that based on future experiment further refinements in our charm hadron decay tables are needed. To conclude, there is still much to do in preparation for the charm hadron era of relativistic heavy ion collisions. 

\section*{\label{sec:acknowledgements}Acknowledgments}
This work has been supported by a grant from the U.S. Department of Energy, grant DE-FG02-04ER41318,
Laboratoire de Physique Th{\' e}orique et Hautes Energies, LPTHE, at University Paris 6 is supported by CNRS as Unit{\' e} Mixte de Recherche, UMR7589. 

\end{document}